\documentclass[journal]{IEEEtran}
\IEEEoverridecommandlockouts

\usepackage{cite}
\usepackage{amsmath,amssymb,amsfonts}
\usepackage{lipsum}
\usepackage{graphicx}
\usepackage{textcomp}
\usepackage{xcolor}
\usepackage{newunicodechar}
\newunicodechar{−}{\ensuremath{-}}
\usepackage{multirow}
\usepackage{breqn}
\usepackage{dirtytalk}
\usepackage{algorithm}
\usepackage{float}
\usepackage{upgreek}
\usepackage{algpseudocode}
\usepackage{booktabs}
\usepackage{subcaption}
\usepackage{multicol}
\usepackage{array}

\usepackage{array}
\usepackage{soul}
\title{Cybersecurity Issues in Local Energy Markets}

\author{
    Al Hussein Dabashi,
    Sajjad Maleki,
    Biswarup Mukherjee, Gregory Epiphaniou, Carsten Maple, 
    Charalambos Konstantinou, Subhash Lakshminarayana 
    
    \thanks{A. Dabashi and S. Lakshimnarayana are with the School of Engineering, University of Warwick, CV47AL, UK. S. Maleki is with the School of Engineering, University of Warwick, CV47AL, UK and ETIS UMR 8051, CY Cergy Paris Universit\'e, ENSEA, CNRS, F-95000, Cergy, France. B. Mukherjee and C. Konstantinou are with the CEMSE Division, King Abdullah University of Science and Technology (KAUST). G. Epiphaniou and C. Maple are with the WMG, University of Warwick. \\
    \emph{Corresponding author:} S. Lakshminarayana,  subhash.lakshminarayana@warwick.ac.uk)}
}

\begin{document}
\maketitle

\begin{abstract}
Local Energy Markets (LEMs), though pivotal to the energy transition, face growing cybersecurity threats due to their reliance on smart grid communication standards and vulnerable Internet-of-Things (IoT)-enabled devices.
This is a critical issue because such vulnerabilities can be exploited to manipulate market operations, compromise participants' privacy, and destabilize power distribution networks.
This work maps LEM communication flows to existing standards, highlights potential impacts of key identified vulnerabilities, and simulates cyberattack scenarios on a privacy-preserving LEM model to assess their impacts.
Findings reveal how attackers could distort pricing and demand patterns. We finally present recommendations for researchers, industry developers, policymakers, and LEM stakeholders to secure future LEM deployments.

% Local Energy Markets (LEMs) are emerging as a powerful solution to the energy transition, with benefits proven by real pilot projects.
% However, LEMs are susceptible to cybersecurity threats, depending heavily on evolving digital infrastructure for reliable operation.
% This includes high-volume communications and numerous Internet of Things (IoT)-enabled devices, whose cybersecurity implications within coordinated, market-driven frameworks remain underexplored.
% To bridge this gap, this article provides an accessible explanation of LEMs, then delves into the communications and devices involved, their associated vulnerabilities, and impacts of exploits.
% As a case study, two simulated cybersecurity attacks on a privacy-preserving LEM model are presented and analyzed, with key insights indicating that cyber weaknesses in LEMs pose significant technical and financial threats, for instance by granting attackers indirect control over energy usage patterns.
% Appropriately, this article provides recommendations for researchers, industry and government, and concludes with implementable guidance for LEM stakeholders to help mitigate cybersecurity issues in LEMs.
\end{abstract}


\section{Introduction}

\subsection{Overview and Motivation}
\IEEEPARstart{A}{midst} the global trend in decarbonization targets, the energy sector is undergoing unprecedented transformation.
Energy markets, in particular, are undergoing rapid structural reformations to accommodate emerging business frameworks driven by the increasing integration of distributed and renewable energy sources (DERs/RES).
This integration, which must be accomplished alongside societies' increasing power demand, causes heavy strain on existing electric transmission and distribution networks.
In response, many solutions are being proposed from the demand-side perspective.

\begin{figure}
     \centering
     \includegraphics[width=1\linewidth]{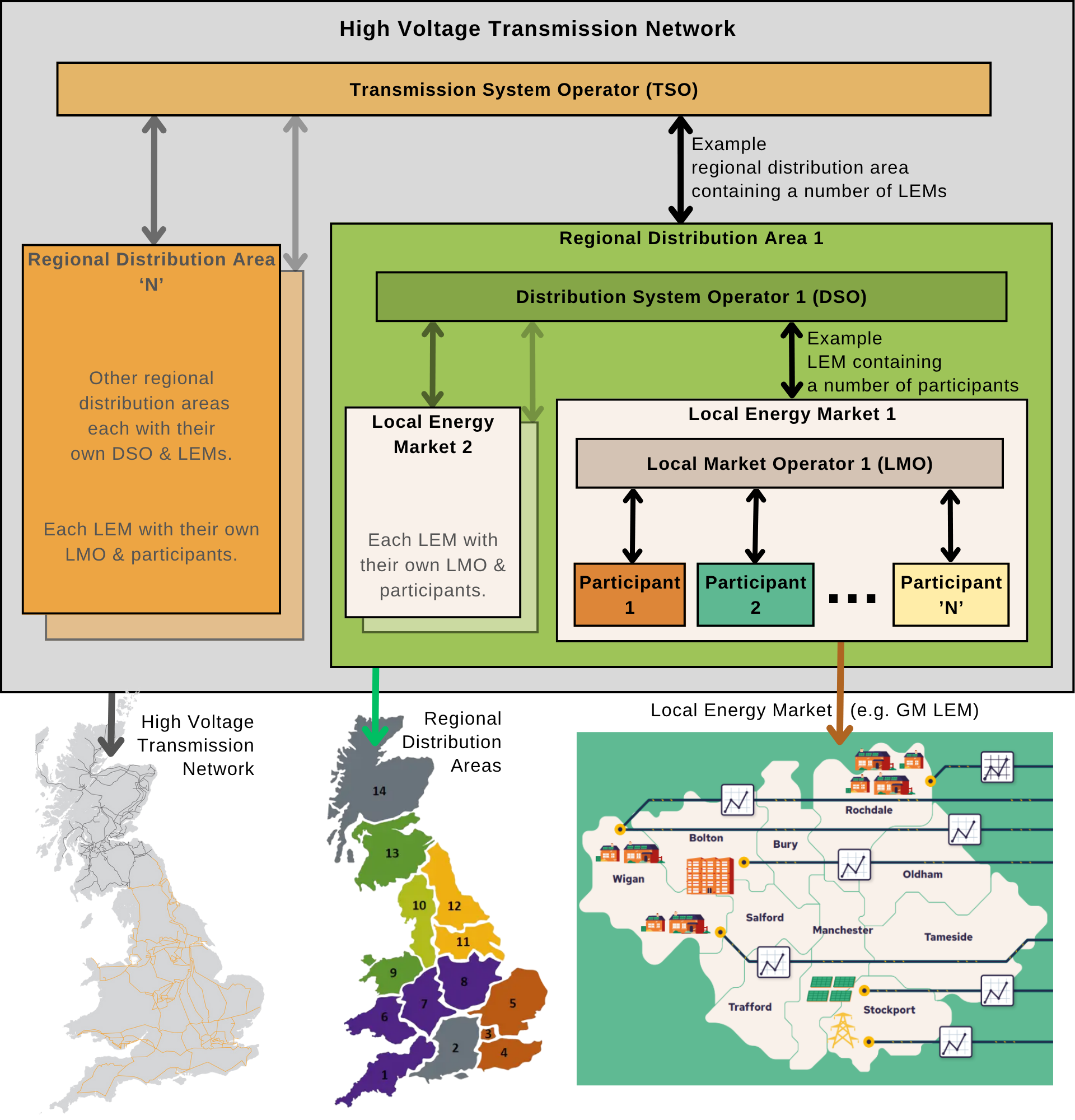}
     \caption{\small{LEM placement within distribution areas and power grid. Multiple LEMs per regional distribution area possible. GB examples illustrate scale, LEM example from Greater Manchester LEM detailed design project (Innovate UK). }}
     \label{fig:lem-overview}
\end{figure}

Among those solutions, local energy markets (LEMs), which are decentralized distribution-level energy trading platforms, are gaining attention as a way to increase the uptake of DERs and RES, allow residential prosumers to sell excess generation for profit, and enhance grid efficiency by matching local generation with local demand first \cite{faia2024local}. Figure \ref{fig:lem-overview} shows a conceptual placement of LEMs within distribution areas and the wider power grid, with examples given from a Great Britain (GB) context to illustrate their relationship and geographical size.
Note how multiple LEMs can operate within a single regional distribution area, enabling for broader participation, thereby expanding the reach of their benefits. %\cite{savvopoulos2021coordinet} cutting refs %\cite{centrica}.
Real pilot trials have proven these benefits; the Cornwall LEM in GB, lead by Centrica (a British international energy and services company) during 2017-2020, saw 310MWh of power traded successfully between over 200 homes and businesses, with greenhouse gas savings of nearly 10,000 tonnes a year. The CoordiNet project in Europe, funded by EU’s Horizon 2020 during 2019-2022, saw pilots in Spain, Sweden, and Greece, and realized benefits such as improved grid coordination and energy trading access for consumers. 

However, with these benefits comes the challenge of cybersecurity. In addition to LEMs' dependence on an  interconnected cyber-physical infrastructure, LEM's operation at the grid's `edge' (composed of high-volume real-time communications with numerous devices from many distributed participants) exposes them to more attack entry points, rendering them arguably more vulnerable than conventional wholesale energy markets (WEMs) \cite{faia2024local}. % Cutting due to references limit of 15: \cite{Chen2025Cybersecurity, Abdelkader2024Securing}

The relevant Internet-of-Things (IoT)-enabled devices for this discussion are high-wattage devices, whose aggregated operation significantly impacts energy consumption patterns. This paper focuses on two main categories: DERs, such as photovoltaic (PV) panels, and energy smart appliances (ESAs), such as electric vehicle (EV) chargers, smart heat pumps, and smart washing machines. Hereafter, we refer to these collectively as ``IoT devices," with this specific context implied.
A common vulnerability found in both categories are weak authentication, such as default user credentials, which could allow adversaries easy access to orchestrate large-scale demand fluctuations by controlling many ESAs \cite{lakshminarayana2025threats}.
Furthermore, for communication standards such as OpenADR and IEC 2030.5, protocol implementation vulnerabilities have been shown, which could be exploited to send false device control commands \cite{Zografopoulos2023Distributed}. 
These cybersecurity and privacy concerns with weaknesses serving as attack vectors ready to be exploited, and the need to encourage a proactive rather than reactive response to them, form the motivation for this article.

\subsection{Related Works and Our Contributions}

The literature on LEM cybersecurity remains limited and often lacks specificity. Mustafa et al \cite{Mustafa2016ALocal} and Dong et al \cite{Dong2022Cybersecurity} both address cybersecurity concerns in LEMs by identifying key threats and standard security requirements, such as authentication and message verification. However, neither works provide guidance adapted to LEM-specific operations such as real-time pricing or trading mechanisms, nor do they quantify the potential impacts of exploits on LEM operation.

%\subsection{LEM-specific gaps}
Dedrick et al \cite{Dedrick2023Assessing} address the aforementioned gaps by providing a quantitative cybersecurity attack analysis showing how LEMs are vulnerable even to unsophisticated low-rate DoS attacks. They also detail LEM specifics, such as message sequences and market clearing, however their study does not include other important cyberattacks such as FDIAs.

Cyberattacks in transactive energy markets (TEMs) are detailed by Dasgupta et al \cite{Dasgupta2021Cyber}, classifying the attacks into three categories. They discuss the adversary roles and what benefits can be attained through each attack. Barreto et al \cite{Barreto2020Attacking} show how an adversary can manipulate customer devices that participate in electricity markets. The manipulation involves FDIA on bids, which change the market's equilibria to the adversary's advantage. Although similar concepts are discussed, neither work focuses on LEMs. 

A significant limitation observed across the reviewed literature is the failure to map LEM communication signals to existing standards, which this article addresses. This is a critical gap, particularly as LEMs move from concept to real-world deployment. Identifying the standards in use is essential for assessing vulnerabilities and anticipating the consequences of potential cyber exploits, especially given the rising frequency of cyberattacks targeting the energy sector. With this foundation, effective prevention, detection, and mitigation strategies, along with robust cyber risk assessments, can be developed. Dong et al \cite{Dong2022Cybersecurity} do suggest the IEEE 2144.1 standard for peer-to-peer (P2P) trading, however P2P is not LEM specific, representing a transaction mechanism rather than a whole-market design such as LEMs. Also Kirpes et al \cite{Kirpes2019Design} in 2019, do suggest IEC 61850 and IEC 62325 for smart meter and energy orders respectively, however these suggestions were mentioned briefly, without further justification, analysis, or mapping to the wider set of LEM communication signals. %such as transaction verification or bidding preferences.

%A significant limitation that appears commonly across the reviewed literature is the failure to map LEM communication signals to existing standards. Dong et al. \cite{Dong2022Cybersecurity} do suggest the IEEE 2144.1 standard for peer-to-peer (P2P) trading, however P2P is not LEM specific, representing a transaction mechanism rather than a whole-market design such as LEMs. Moreover, none of the studies analyze the vulnerabilities of standards or assess the potential impact on LEM operations if such vulnerabilities were exploited.

Accordingly, this article makes the following contributions:
\begin{enumerate}
    \item An accessible explanation of LEMs, including key participants, entities, and their communication flows, while identifying the specific attack vectors that may be targeted by insider or external attackers, in both clearly-drawn figures and in-text. 
    \item A detailed mapping of LEM communication signals to existing standards, such as OpenADR and IEC 2030.5, along with descriptions, justifications, associated vulnerabilities, and the potential impacts of exploits on LEM operation. In addition, key vulnerabilities of IoT devices are examined, along with their implications on LMO system security and participant privacy.
    \item Two simulated FDIA case studies demonstrating the attack methodology and its technical and financial impact on LEM operation, complemented by a summary of key findings and related studies (see Table \ref{tab:key_findings_LEM}), followed by short-term and long-term recommendations for researchers, industry, and policymakers, then concluding with actionable guidance for LEM stakeholders.
\end{enumerate}

%\sub{Remove this and refer to the table separately.} \Alh{Addressed}
%\Alh{Although in our case the increase is in the reported optimal demand, see Table \ref{tab:key_findings_LEM} to see how this is similar to how Shekari et al \cite{shekari2021mamiot} showed that by increasing the power demand of compromised IoT devices, real-time pricing can be increased intentionally, which in turn benefits power generation utilities. Their study focused on the manipulation of market via IoT (MaMIoT) using load-altering attacks (LAAs) \cite{maleki2025survey}. Findings from Ospina et al \cite{ospina2023feasibility} similarly reveal how LAAs at specific loads can substantially raise electricity prices in LEMs.}

% \harrys{it is unclear how the literature gap is being addressed in this paper - we do not aim to do that. This is not a research paper; we need to position these works that they did something, however, many more things need to be done, thus we lay out the foundation principles for one to consider studying this topic } \Alh{Addressed, I have updated the literature review and contribution of this paper above.}

\begingroup
\renewcommand{\arraystretch}{1.7}
\begin{table*}[htbp]
    \footnotesize
    \centering
        \caption{Summary of key findings of works investigating different types of cybersecurity attacks on LEM operation}
    \begin{tabular}{>{\raggedright\arraybackslash}p{1.5cm}  % Authors
                    >{\raggedright\arraybackslash}p{1.7cm}  % Cyberattack
                    >{\raggedright\arraybackslash}p{11cm}  % Key Findings
                    >{\raggedright\arraybackslash}p{2cm}}  % Year and Reference
    \hline
    \textbf{Authors} & \textbf{Cyberattack} & \textbf{Key LEM-relevant Findings} & \textbf{Year, Reference} \\
    \hline
    Andriopoulos \newline et al. & FDIA & 1. IoT-based LEMs show significant vulnerability to FDIAs on smart meters, HVAC systems, and photovoltaics, causing voltage violations and increased operational costs.
    \newline 2. A nodal DLMP-based LEM structure effectively mitigates these risks by enhancing grid visibility and facilitating early attack detection via nodal pricing. & 2025, \cite{Andriopoulos2025Cyber} \\

    Ospina \newline et al. & LAA & 1. Coordinated LAAs targeting IoT-connected devices at distribution levels could significantly manipulate LMPs within LEMs.
    \newline 2. Such LAAs could facilitate energy storage arbitrage, enabling attackers to exploit manipulated LEM conditions for substantial financial gains. & 2023, \cite{ospina2023feasibility} \\

    Dedrick \newline et al. & DoS & 1. LEMs are highly vulnerable to low-rate DoS attacks, with just 5–10\% message loss causing notable market disruptions.
    \newline 2. Random bid losses from DoS attacks distort price distributions, subtly biasing LEM outcomes and affecting power flows without immediate detection. & 2023, \cite{Dedrick2023Assessing} \\

    %(Cutting due to references limit, also this work focuses on wholesale energy markets, not LEMs:) Shekari \newline et al. & MaMIoT & 1. MaMIoT attacks, leveraging high-wattage IoT botnets, could predictably alter electricity demand to manipulate prices in LEMs, benefiting attackers financially.
    % \newline 2. Such IoT-driven manipulations pose feasible and serious financial threats to targeted LEM participants and market stability. & 2021, \cite{shekari2021mamiot} \\

    %(cutting due to reference limit of 15 + this is old reference from 2014) Barreto \newline et al. & FDIA-induced LAA & 1. Integrity attacks on demand response systems within smart grids could significantly affect LEM operations, causing sudden overloads and operational instability.
    %\newline 2. Strategic adversaries could exploit control signals in LEM-related demand response programs to fraudulently benefit from manipulated market conditions. & 2014, \cite{barreto2014cps} \\

    This work's \newline case study & FDIA & 1. Manipulated DLMP signals by adversarial prosumers at critical network nodes can significantly distort local pricing in LEMs, influencing market settlement outcomes.
    \newline 2. Coordinated DLMP manipulation can overload the system during low-preparedness hours, causing network congestion, operational cost spikes, and inflated electricity prices for consumers in LEMs.
 & 2025 \\

    \hline
    \end{tabular}
    \label{tab:key_findings_LEM}
\end{table*}
\endgroup

\section{Technical Background of LEMs}

This section explains the mechanisms and entities involved in LEMs, including the data contents of communication signals and the importance of signal integrity for secure operation.

%cutting due to word limit, but keep just in case:
%LEMs are distinct energy trading platforms characterized by their market-based structure, local geographical boundaries, and close integration with the distribution grid. They differ from peer-to-peer (P2P) energy trading, which are direct bilateral transaction mechanisms, rather than full market designs, and do not necessarily consider electrical proximity; from virtual power plants (VPPs), which aggregate RES from geographically broader areas for commercial purposes; and from energy communities, which prioritize cooperative ownership without requiring explicit market mechanisms, although LEMs can support community-based initiatives.
% \subsection{Distinguishing LEMs from Similar Concepts}

\subsection{LEM Entities and their Responsibilities}
As can be observed from Fig. \ref{fig:lem-signals}, there are four main entities concerning LEMs; the Transmission System Operator (TSO), the Distribution System Operators (DSOs), the local market operator (LMO) and the participants. At the operational core is the LMO, responsible for managing financial transactions through market-clearing schemes. The LMO aggregates generation offers, consumption, and flexibility bids from local participants. DSOs handle technical operations of the distribution grid, supplying real-time grid data and local flexibility requirements. TSOs manage signals related to the WEM, such as wholesale market prices and accepted bids. Figure \ref{fig:lem-signals} illustrates these relationships, highlighting the specific entities involved in LEM transactions and their respective roles.

\subsection{Communication Signals between LEM Entities}
%\sub{Add more information in this subsection as discussed, increase the length by atleast 5-6 sentences.}\Alh{Addressed:}

\begin{figure}
    \centering
    \includegraphics[width=1\linewidth]{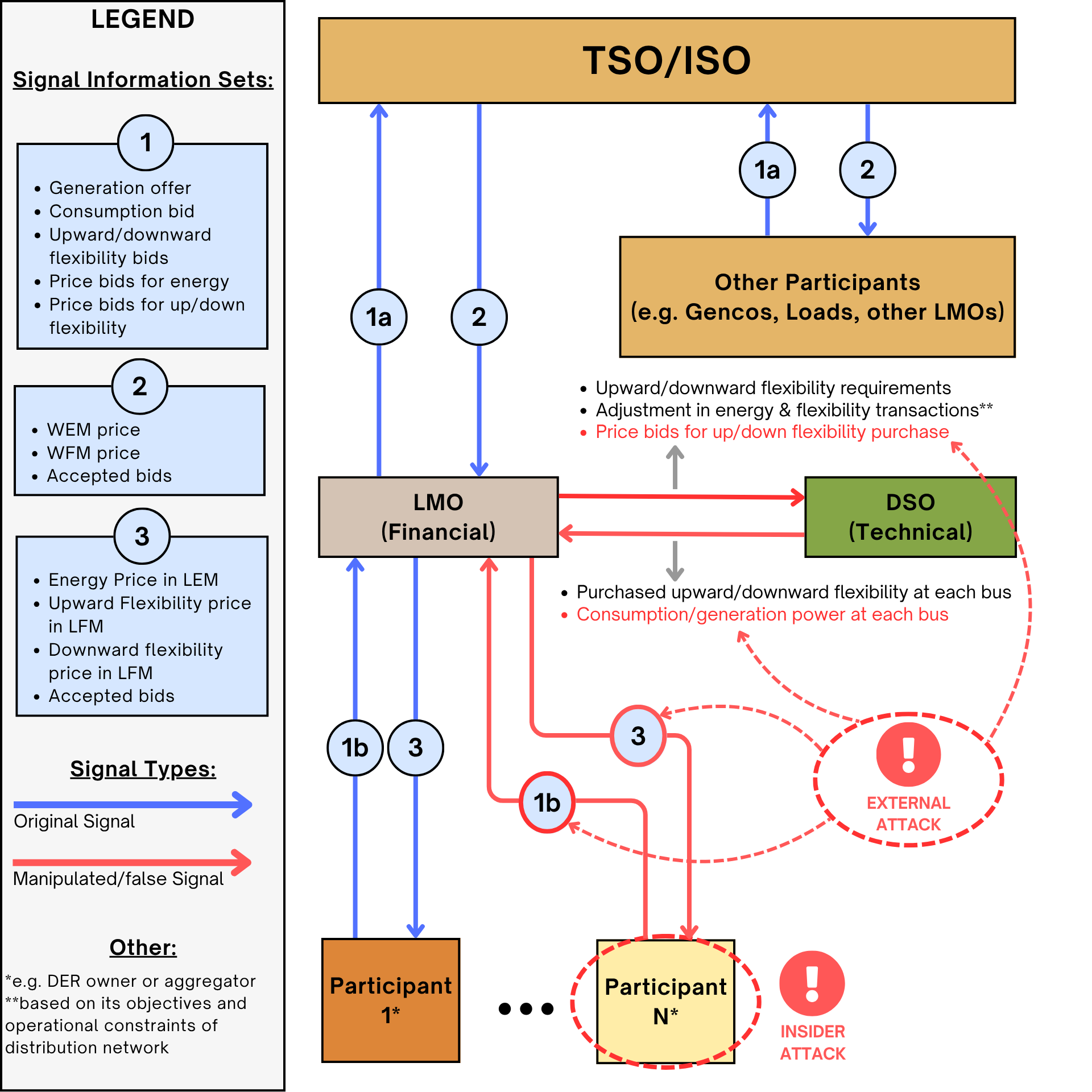}
    \caption{\small{Signals transmitted between LEM and other power grid entities: Signal 1a: Represents upstream request from LMO to TSO, e.g. generation bids for procuring required energy from WEM. Signal 2: Upstream response from TSO, e.g. accepted bids. Signal 1b: Downstream communications from participant to LMO, e.g. generation offer. Signal 3: downstream communication from LMO to participant, e.g. energy price (DLMP). Insider attack: manipulation of data before transmission. External attack: compromising signal in transmission, then modifying.}}
    \label{fig:lem-signals}
\end{figure}

Effective operation of LEMs depends on seamless data exchange among its entities. As depicted in Fig. \ref{fig:lem-signals}, these interactions comprise three main signal information sets. Signal set 1 is the conceptual starting point for LEM operation, because it is only transmitted from the smaller entity to the larger entity such as from the LEM participant to the LMO (signal 1b) and from the LMO to the TSO (signal 1a). This is so the larger entity can collate all information from the other smaller entities, and communicate a response with information that the smaller entities need. Signal set 1 contains information regarding the: 1) generation offers and consumption bids 2) upward/downward flexibility bids 3) price bids for energy and 4) price bids for upward/downward flexibility bids. The response signal from the larger entity are signal sets 2 and 3, sent from the TSO to LMOs and from LMOs to their participants. Signal set 2 transmits wholesale energy and flexibility market prices, and the bids accepted at the wholesale level. Signal set 3 communicates LEM-specific energy and flexibility prices along with accepted local bids. Note that the energy price in LEM is the same as the distribution locational marginal price (DLMP), which is explained in section \ref{case-study-section}. These signals ensure proper LEM operation, hence maintaining their integrity is crucial.

\subsection{Importance of Signal Integrity and its Susceptibility to Attacks}
Integrity of communication signals within LEMs is critical to ensure accurate market clearing, fair price formation, and reliable grid operations. Any compromise to signal integrity, whether through inside or external cyberattacks, such as malicious prosumers falsifying bids, could significantly distort market outcomes. As indicated by the red lines in Fig. \ref{fig:lem-signals}, attackers may introduce false signals, misleading LMOs and DSOs, disrupting financial settlements, and causing potential operational instability. Vulnerabilities such as these emphasize the necessity of securing communication standards and protocols, which the next section addresses by discussing established protocols, their known vulnerabilities, and their potential impacts if exploited.

%\sub{Put more info in the figure heading as discussed.} \Alh{Addressed.} 

\begingroup
\renewcommand{\arraystretch}{1.7}
\begin{table*}
    \scriptsize
    \centering
    \caption{Vulnerabilities of LEM-relevant communication protocols and their impacts if exploited. LEM Signal column refers to the corresponding proposed communication standard for the LEM signals shown in Fig. \ref{fig:lem-signals}.}
    \begin{tabular}{>{\raggedright\arraybackslash}p{1.5cm}  % Protocol Name
                    >{\raggedright\arraybackslash}p{1.5cm}  % Signal in LEMs
                    >{\raggedright\arraybackslash}p{4cm}  % Description
                    >{\raggedright\arraybackslash}p{3cm}  % Vulnerabilities
                    >{\raggedright\arraybackslash}p{6cm}}  % Impact
        \hline
        \textbf{Protocol} & \textbf{LEM Signal} & \textbf{Description} &\textbf{Vulnerabilities} & \textbf{Impact if Exploited} \\
        \hline
        
        UFTP \newline (USEF Flex Trading Protocol) & Signal 1b \newline Signal 3
        & Protocol for market interactions \& flexibility trading between aggregators \& DSOs to resolve grid constraints by applying congestion management or grid-capacity management.
        & No publicly known vulnerabilities to date. Developed with privacy and security by design principles, such as data minimization, role-based access control, secure message transmission.
        & Attackers could manipulate flexibility bids, spoof or block FlexRequests \& FlexOffers, or tamper with settlement processes. This could lead to incorrect activation of flexibility, grid imbalances, financial losses, market manipulation, \& erosion of trust between aggregators, DSOs, \& prosumers.   \\

        OpenADR \newline (Automated Demand Response) \newline \cite{opengridsystems2023flexibility} & Signal 1b \newline Signal 3
        & {Open standard for automated DR proposed for exchanging dynamic price \& reliability signals between LEM participants and LMOs.} \ %for automated DR, used to exchange dynamic price \& reliability signals amongst utilities, TSOs, \& energy management \& control systems. \sub{tso??}
        & {OpenADR 2.0 has possible implementation vulnerabilities (e.g. in CERT Java) \cite{Zografopoulos2023Distributed}.}
        &  {Exploit of transaction vulnerability could lead to falsified bids or price signals in LEMs, causing unfair market outcomes, financial losses, and grid imbalances. LMOs may process incorrect data, undermining trust, disrupting flexibility dispatch, and compromising both economic integrity and grid stability.}    \\
        
        IEEE 2030.5 \newline (Smart Energy Profile 2.0) \newline \cite{opengridsystems2023flexibility} & Signal 1b \newline Signal 3
        & Enables utility management of the end-user energy environment, such as DR, load control, time-of-day pricing, distributed generation management, \& electric vehicle management.
        & Limited publicly known vulnerabilities, however, vulnerabilities are possible via improper implementation of the protocol.
        & Attackers could issue false DER commands/suppress legitimate control signals, e.g. curtailing generation at incorrect times, leading to local supply \& demand imbalances \& manipulated energy prices. In worse cases, coordinated false commands could destabilize the microgrid. Privacy is also a concern; a breach could expose participants' unique personally-identifiable usage patterns.  \\

        IEC 61850 \newline \cite{opengridsystems2023flexibility}  & Signal 1b
        & IED* communications standard at digital substations. {Proposed for sending energy demand \& excess energy from LEM participants' smart meters to LMO \cite{Kirpes2019Design, opengridsystems2023flexibility}.} 
        & e.g. CVE-2022-3353, which affects multiple Hitachi Energy products. CVE-2024-12169, EUVD-2025-19012
        & An attacker could exploit the vulnerability to force the IEC 61850 MMS-server communication stack to block new MMS-client connections through a specially crafted message sequence.  \\

        IEC 62325 \newline (MADES: MArket Data Exchange Standard)  & Signal 1a \newline Signal 1b \newline Signal 2 \newline Signal 3
        & Set of energy market communication standards {proposed for conveying order information, such as buy and sell orders, between participants and LMOs \cite{Kirpes2019Design}.}
        & {No publicly known vulnerabilities. Built with IEC 62325-503 ``Cybersecurity for the Energy Market" with security features such as encrypted channels.}
        %e.g. CVE-2024-12169, which allows for the forced restart of an affected Communication Module Unit (CMU) in RTU500 systems (Hitachi Energy) \cite{NVD2024CVE}. 
        & Attackers could manipulate core market exchanges (e.g. scheduling, settlements, capacity nominations) which may compromise message integrity or non-repudiation despite encryption, causing inaccurate pricing, disrupted WEM–LEM coordination, financial imbalances, and decreased trust in European-style LEM integration. \\
    \hline
    \multicolumn{5}{@{}p{\textwidth}@{}}{\scriptsize {*Intelligent Electronic Device.}}     
    \end{tabular}
    \label{tab:protocol-vulnerabilities}
\end{table*}
\endgroup

%PArt of table previously: cutting due to reference limit of 15:\cite{franzl2021technical}} \\
% This vulnerability only applies when secure communication via IEC 62351-3 (TLS) is enabled. This critical flaw in RTU500 systems could cause potential DoS \& subsequent disruptions when an attacker triggers a specific sequence to restart the CMU. Impact industrial DSO's control systems reliant on RTU500, causing potential downtime of LEMs \cite{Ghanem2020Bandwidth}.  \\ 
%DNP3 \sub{delete?} \newline(IEEE 1815) & Signal 1b
%& Set of communication protocols primarily used in process automation systems, used in utilities such as power \& water. 
%& Eight vulnerabilities have been identified \& implemented by \cite{Vasiliki2022Risk} exposing deep vulnerabilities-by-design issues.
%& The DNP3-centred simulated cyberattacks enables attackers the possibility to disrupt the operation of future LEM implementations \& damage critical equipment. 

\section{Cyber Vulnerabilities in LEMs}

After discussing the importance of secure communications previously, this section focuses on the vulnerabilities of  proposed communication standards and of IoT devices.

\subsection{Vulnerabilities in Communication Standards}

There are no specifically agreed upon standards for LEM signals, hence various implemented local and flexibility market platforms resort to proprietary methods, as shown in the 'Flexibility Market Report' published by Open Grid Systems in 2023 \cite{opengridsystems2023flexibility}.
Proprietary methods, which are usually not open source, raise significant cybersecurity concerns, since they can not be verified nor contributed towards by a community of experts.
They should therefore require security verification by regulators, such as Ofgem in the UK.
In this section, the focus is on proposing existing standards due to: 1) the current large efforts being placed on standardizing power system and supporting services data (e.g. IEC TC 57) and 2) the ready availability and widespread deployment for most of these standards. Furthermore, these proposed standards are relevant to LEMs due to their capability of encapsulating market information as part of their data domains \cite{opengridsystems2023flexibility}.  % keep commented here for reference, e.g. may get contacted about this point in the future by researchers: https://www.energynetworks.org/newsroom/ena-elexon-and-openadr-alliance-sign-letter-of-intent-to-pave-the-way-for-international-flexibility-standard, and: https://tc57.iec.ch/

\subsubsection{The USEF Flex Trading Protocol (UFTP)} is developed by the Universal Smart Energy Framework (USEF) foundation and is designed for communicating flexibility data between aggregators and DSOs specifically, however it can be used as a stand-alone protocol for carrying market data such as consumer bids and generation offers. % \cite{uftp} cutting refs
Hence, we deem this protocol suitable to transmit signal set 1b and 3 shown in Fig. \ref{fig:lem-signals}. 
Regarding security, the UFTP has no publicly documented vulnerabilities to date and it was developed with strong privacy and secure-by-design principles, such as data minimization and secure message transmission via Transport Layer Security (TLS). TLS is a cryptographic method to provide end-to-end security of data sent over the Internet. This highlights how cybersecurity measures in protocol-development can be prioritized from the outset, setting a valuable precedent for future LEM communication standards. Nevertheless, if vulnerabilities emerge or improper implementations occur, attackers could falsify or block various elements of the protocol, such as its  FlexRequests and FlexOffers, disrupting the flexibility market and creating local grid imbalances, financial losses, and eroding trust between LEM stakeholders. UFTP has been used in practice for Project FUSION, by Scottish Power Energy Networks during 2018–2023 in the UK, which trialed the trading of commoditized local demand-side flexibility in a structured and competitive market.
% \cite{versmissen2020project} cutting refs

\subsubsection{The OpenADR protocol \cite{opengridsystems2023flexibility}} which is also proposed for signal set 1b and 3, is developed by the Open Automated Demand Response Alliance to standardize and simplify demand response (DR) and DER management for utilities and aggregators, while enabling consumers to control their energy usage. %\cite{openadr} cutting refs
Earlier works have reported that OpenADR contains over 700 implementation vulnerabilities in the open-source CERT Java library \cite{Zografopoulos2023Distributed} (which is not necessarily from the protocol’s specification itself). This highlights that improper implementation is likely to create weaknesses that could be exploited, permitting remote attackers to access sensitive data such as usernames, system properties, and installation directories \cite{Zografopoulos2023Distributed}. Furthermore, the newer OpenADR 3.0 specification, released in late 2023, may have addressed several of these concerns, for example in this newer version, widely accepted security standards are adopted instead of custom requirements. %( cutting due to references limit of 15:) Also, mutual TLS is no longer required; only the server provides a certificate, thereby simplifying the setup and configuration process. The requirements of TLS have also been revised to reduce complexity while maintaining a secure communication channel \cite{grabarz2025convention}. 

\subsubsection{The IEEE 2030.5-2023 standard \cite{opengridsystems2023flexibility}} is also known as SEP2 (Smart Energy Profile 2) and is intended for DER management and demand response communications. This standard is proposed for signal sets 1b and 3. The standard is used extensively in regions prioritizing RES integration and smart grid development. It is a mandated standard for regulatory compliance in areas such as California, and it has also gained traction in other parts of the US, Canada, and Australia. The IEEE 2030.5-2023 standard employs robust encryption and authentication, however, implementation vulnerabilities could still lead to opening attack vectors such as unauthorized DER control commands, false event signals, and data breaches. Such scenarios may lead to manipulated market prices, privacy violations, or interruptions in local demand-supply balance in LEMs, directly harming market integrity and distribution grid reliability.

\subsubsection{The IEC 61850 standard \cite{opengridsystems2023flexibility}} is a widely deployed international standard for communication and system architecture in digital substations and  distribution networks. In this work, the IEC 61850 is proposed for smart meter communication of prosumer supply and bid data (signal 1b). It has known vulnerabilities such as CVE-2022-3353, which specifically affects the MMS-server. Attackers could exploit this by sending specially crafted message sequences to block legitimate MMS-client connections, thereby interrupting critical LEM signals. Consequently, this could disrupt bid submission and flexibility activations, leading to market mismanagement and distribution-level grid issues. Recent examples include EUVD-2025-19012, which is vulnerable to Man-in-the-Middle (MITM) attacks due to improper TLS certificate validation. While standards to secure these protocols exist (the IEC 62351 series defines security measures for IEC 61850 and other standards), they are not always implemented.

\subsubsection{The IEC 62325 standard} is also known as MADES (MArket Data Exchange Standard), and is a set of energy market communication standards proposed for signal sets 1a, 1b, 2, and 3 since it facilitates for the exchange of order information, such as buy and sell orders \cite{Kirpes2019Design}. It incorporates IEC 62325-503 cybersecurity provisions, including encrypted communication channels. While there are no publicly known vulnerabilities, potential impacts of exploits include adversaries manipulating core market functions such as scheduling, settlements, or capacity nominations, which may compromise message integrity or non-repudiation. This could lead to inaccurate pricing, disrupted coordination between wholesale and local markets, financial imbalances, and reduced trust in European-style LEM integration. %cutting due to reference limit of 15: \cite{franzl2021technical}.
% \cite{IEC62325} cutting refs

See Table \ref{tab:protocol-vulnerabilities} which tabulates the aforementioned communication standards, their mapping to the signal information sets in Fig. \ref{fig:lem-signals}, their vulnerabilities and potential impacts if exploited.

\subsection{Vulnerabilities in IoT Devices}

Vulnerabilities in IoT devices also pose significant threats to the secure operation of LEMs, especially when coordinated intentionally, weaknesses in such devices could allow adversaries to alter the devices' energy usage or orchestrate simultaneous switching events, thereby disrupting expected load patterns. This type of disruption, is referred to as a load-altering attack (LAA) \cite{maleki2025survey}, and can lead to substantial deviations from the LMO's anticipated forecasts, with direct consequences on market interactions, bid acceptance, and ultimately, LEM clearing prices. A further cyber threat involves unauthorized access to LMO platforms to obtain participant identities, raising significant privacy concerns. 

Examples of relevant vulnerabilities in this context includes weak authentication \cite{lakshminarayana2025threats}, in which a common vulnerability is the default user credential of a DER's or ESA's device management portal. For example, solar inverters manufactured by  Deye, a large Chinese solar inverter company operating in the global market, use ``admin" as default for both the username and password. This weakness could be exploited by adversaries for easy access. % \sub{where does this company operate? write something like ``a large-inverter manufacturer in US...''}\Alh{Addressed}
Remote Code Execution (RCE): Describes when an adversary who has gained access can run malicious programs on DERs or ESAs. This could before example to disrupt the devices' functioning, to allow remote control or to extract sensitive information. For EV chargers, which continue to proliferate rapidly across society, they have become increasingly susceptible as evidenced by CVE-2025-5747, CVE-2025-5748, and CVE-2025-5828.
The above vulnerabilities are the most relevant to LAAs which could affect LEMs, however other vulnerabilities in IoT devices exist such as vulnerable communication mediums  and insecure web interfaces \cite{lakshminarayana2025threats}.

\section{Case Study of Cyberattack Simulation on LEM} \label{case-study-section}

This section presents a case study to illustrate the negative impacts of cyberattacks on LEM operations. Two FDIA scenarios are simulated, one conducted by an insider LEM participant and another by an external malicious actor, highlighting their effects on electricity prices (\$/kWh) and power demand. The electricity prices are represented by DLMPs, which indicate the localized cost of supplying electricity at specific nodes of the distribution grid, including factors such as network congestion and losses.

%  (they are calculated through solving a distributed optimization problem as described in the foll)

\subsection{Overview of the LEM Model and Data Privacy}
This case study LEM model consists of multiple participants and an LMO, similar to Fig. \ref{fig:lem-signals}. The LEM participants send their optimal demands and generation to the LMO, representing signal 1b in Fig. \ref{fig:lem-signals}. The LMO collects this information, purchases energy from the WEM accordingly, and computes DLMPs (using the distribution optimization scheme proposed in the next subsection). The LMO then sends these DLMPs back to the participants, representing signal 3 in Fig. \ref{fig:lem-signals}, who then, based on received DLMPs update their demands/generations for the day-ahead to be sent to the LMO and so on. The test case in this section is a modified version of the IEEE 69-bus distribution grid, which has 48 load buses with participants containing Electric Vehicles (EVs), photovoltaics, energy storage units, and flexible loads. %cutting due to reference limit of 15: \cite{savier2007impact}

The information of LEM participants contains sensitive and identifiable data such as LEM participant IDs and EV arrival and departure schedules. To enhance the privacy and maintain confidentiality, we employ distributed optimization for market clearing, as explained in the next subsection.

\subsection{Market Clearing Scheme via Distributed Optimization}

LEM clearance can be solved either as a centralized or distributed optimization problem. Distributed optimization presents a more advantageous mechanism for deriving DLMPs and optimal bids for participants due to the preservation of the confidentiality of participants and the LMO’s private data. This works by each participant independently performing a local optimization based on the DLMPs to determine their optimal load demand for the subsequent day, rather than having the LMO perform a centralized optimization for all participants.

For the specific method of distributed optimization, we employ the alternating direction method of multipliers (ADMM) \cite{kabirifar2025distributed}. ADMM operates by iteratively exchanging locally optimized decisions from participants with the LMO, and in return, the LMO shares signals to guide convergence toward a global optimum. These data exchanges can be attacked by insider or external threat actors, jeopardizing the optimization process and ultimately impacting LEM operation. The two FDIA scenarios are discussed in the next subsection.

\subsection{Description of Cyberattack Scenarios}

Here, we examine two FDIA scenarios on the transmitted data between the participants and LMO, with the compromised signals representing signals 1b and 3 in Fig. \ref{fig:lem-signals}.

\subsubsection{Insider attack} Refers to malicious participants deliberately altering or falsifying the data they send to the LMO, deviating from their actual values to manipulate market outcomes. The main motivations include increasing the profit (for prosumers and aggregators) and savings on energy bills (for consumers). Fig. \ref{fig:lem-signals} shows an illustration of such an attacker (labeled `participant N') within the LEM framework and the compromised signals in red. 

% The impact this FDIA has on the LEM will be discussed in the next subsection. 

% In this case study, we assume that the adversarial prosumers transmit optimal demands that are 30\%-50\% higher than their actual values. Since the effectiveness of the implemented ADMM depends on the accurate reporting of information, the LEM is impacted through subsequent optimizations, each carrying errors from the previous iterations. 

\begin{table}[b]
    \centering
    \caption{Case study environment and parameters}
    \begin{tabular}{|>{\centering\arraybackslash}p{2.4cm}|
                    >{\centering\arraybackslash}p{5.1cm}|
                                  }
        \hline
        Parameter & Value \\ \hline
        Bus System & IEEE 69 bus \\ \hline
        Load buses & 48 \\ \hline
        Grid type & Distribution \\ \hline
        Voltage level& 12.66 kV \\ \hline
        \multirow{2}{*}{Load types} & EV charge, Flexible load, Conventional fixed load, BESS charge \\ \hline
        Generation types & PV, EV discharge, BESS\\ \hline
        Market type & Day-ahead\\ \hline
        Bidding Period & 24 hours\\ \hline
        Attacker(s) & External/ Malicious market participant(s) \\ \hline
        Attacked signals & Demand(s)/Generation(s) values- DLMPs\\ \hline 
        The most sensitive targets in scen. 1& \multirow{2}{*}{Buses 50, 49, 21}\\ \hline
        The highest demand & Bus 61 \\ \hline
        The highest demand time & \multirow{2}{*}{19 h} \\ \hline
        Penalty parameter of ADMM ($\rho$) & \multirow{2}{*}{0.2}\\ \hline 
        Flexible load share & $5\%$ of total load\\ \hline
        Number of EVs & 215 \\ \hline
        Number of PVs & 221 \\ \hline
        Number of batteries & 217 \\ \hline

    \end{tabular}
    \label{tab:my_label}
\end{table}

\subsubsection{External attack} Refers to a malicious actor outside the LEM framework who manages to intercept transmitted signals and inject an FDIA on the signal's data. Both the signals sent from the participants to the LMO, or vice versa, are susceptible to this attack, as shown in  Fig. \ref{fig:lem-signals} (external attack on signals 1b and 3). In this case study, the malicious actors manage to intercept signal 3 and tamper with the DLMP data. 

\begin{figure}[htbp]
    \centering

    \begin{subfigure}[b]{0.99\linewidth}
        \centering
        \includegraphics[width=\linewidth]{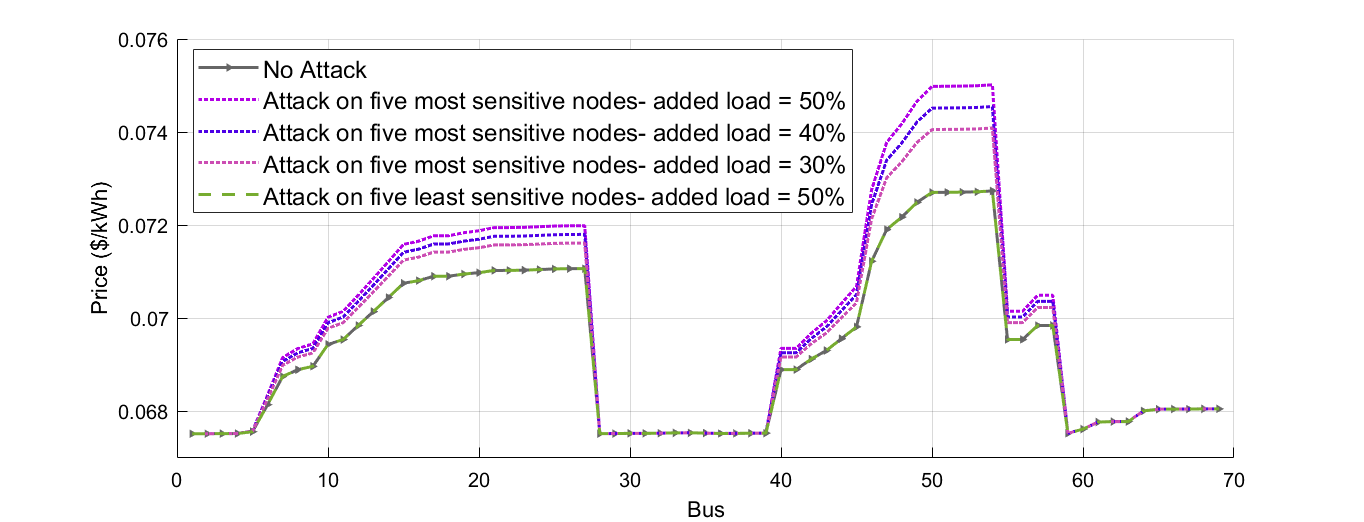}
        \caption{}
        \label{fig:scen1}
    \end{subfigure}

    \begin{subfigure}[b]{0.99\linewidth}
        \centering
        \includegraphics[width=\linewidth]{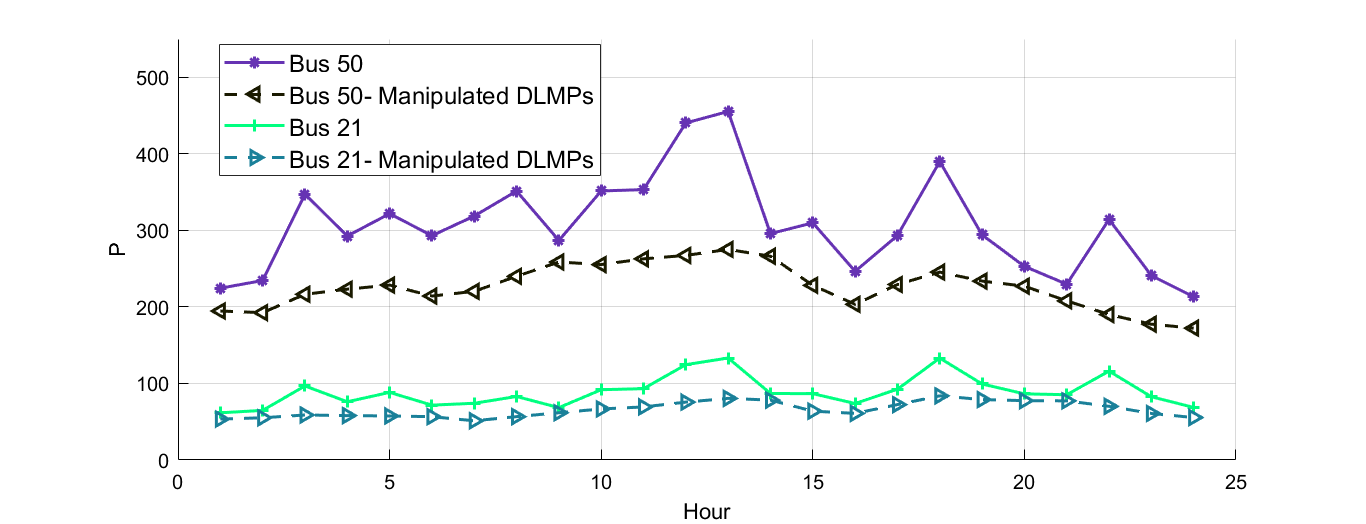}
        \caption{}
        \label{fig:scen2}
    \end{subfigure}
       
    \caption{\small{Two FDIAs: Scenario (a) malicious prosumer insiders (axes: Bus indices and price) and Scenario (b) malicious external actor (axes: hour and demand)}}
    \label{fig:lem-cyberattack}
    \end{figure}

The results on the impact of these attacks are illustrated in Fig. \ref{fig:lem-cyberattack} and explained in the following subsection.

% \begin{figure}
%     \centering
%     \includegraphics[width=1\linewidth]{Figures/graph.png}
%     \caption{Two FDIAs: Scenario 1) malicious prosumer insiders (Axes: Bus indices and price) and Scenario 2) malicious external actor (Axes: hour and demand)}
%     \label{fig:lem-cyberattack}
% \end{figure}

\subsection{Results and Insights}

\subsubsection{Insider Attack} 
{Fig \ref{fig:scen1} shows the corresponding attack impact at 16 hour of the day. In this case study, we assume that the adversarial prosumers transmit optimal demands that are 30\%-50\% higher than their actual values which causes DLMP increase by 1.86\%-3.13\% in the most impacted nodes. As expected, a higher degree of manipulation in reported values leads to a greater deviation of DLMPs from their optimal values.}  The extent of the attack's impact is directly influenced by the pre-attack load level of the manipulated signal, as the deviation of the reported value from the optimal value (real demand) is dependent on the optimal value itself. Importantly, the location of the node(s) under attack also plays a crucial role. Factors such as distance from the root node, energy storage, and local generation significantly affect node sensitivity. In our test case, the attacks on buses 50, 49, and 21 cause the highest impact. 
Thus, it is important for the system operator to consider these factors while deploying defense solutions. 

%\sub{``Fig. XX shows the corresponding attack impact at xxx hour of the day.'' The first sentence must show the impact -  i.e., ``this attack causes X \% increase in the blah blah...''}

% Furthermore, from the figure, we observe that the attack impact critically depends on the location (within the distribution network) at which the attack is injected.

% Note that the reported results in \ref{fig:scen1} illustrate the DLMPs in hour 16.

% Furthermore, experiments conducted in this case study reveal that manipulated prosumer signals at certain nodes have a negligible impact on the final DLMP. 

%This shows that the security of certain participants and communicated signals is more crucial than others, and identifying them can be beneficial for deciding on security investments.

% However, our observations indicate that the load level of the attacked bus is not the sole determinant of a target's sensitivity to attacks. 

\subsubsection{External adversaries} The results show how the manipulated DLMPs affect the changes of power demands at two sample buses (Bus 50 and 21) throughout 24 hours. {These results illustrate that the consumption pattern of customers can be altered via manipulating DLMPs. Note that if the attacker tends to increase the demand during certain hours, they can intentionally decrease the price during that hour, and increase it for the rest of the day to shift all the flexible loads to the cheap hour(s).}
% \sub{Last two sentences are Not very clear...}

\subsubsection{Insights of this case study} In the case study, the attacker performs systematic and sustained manipulation of DLMPs throughout the day using a consistent pattern. Although this alone alters consumption behavior to some extent, a more targeted strategy, hypothetically, would involve adjusting DLMPs only during specific hours. By lowering prices at selected times, the attacker could encourage participants to shift demand into those windows, while raising prices at other times to suppress demand elsewhere. This manipulation would enable the attacker to indirectly control demand patterns, potentially overloading the grid when the LMO is unprepared. Such disruptions risk network congestion, line stress, and increased grid operational costs, leading to inflated actual DLMPs and consumer payments. Another key insight is that vulnerabilities within distributed market optimization systems, if exploited by internal or external adversaries, pose technical and economic risks to network LMOs, potentially leading to higher electricity costs for consumers.

The case study also highlights malicious insider prosumers as significant threats, capable of discreetly falsifying data submissions to influence market outcomes. This emphasizes the critical need for enhanced internal monitoring, robust authentication processes, and well-defined participant guidelines to maintain market integrity, challenges and actionable recommendations for which are further explored in Sections VI and VII.

\subsection{Case Study Assumptions and Future Work}

Although Fig. \ref{fig:lem-signals} illustrates interactions between the LMO and TSO for WEM participation, our case study focuses solely on internal LEM dynamics. As such, WEM-related influences, such as price feedback loops or bulk energy procurement, are not yet incorporated however offer a valuable direction for future work. Additionally, the attack scenarios assume static, predefined adversarial behavior. Future studies could model adaptive attackers capable of responding to system feedback, employing stealthier techniques, or executing coordinated multi-vector attacks.

%\subsubsection{WEM–LEM interactions} Although Fig. 2 illustrates interaction between the LMO and TSO for participating in the WEM, our case study focuses solely on LEM dynamics. Consequently, potential impacts from WEM-related signals on the LEM, such as price feedback loops or bulk energy procurement, are not considered.

%\subsubsection{Static adversarial behavior} The attack strategies applied in both scenarios are static and predefined. In real settings, adversaries may adapt their behavior based on system feedback, employ more stealthy techniques, or launch combined multi-vector attacks, which are not represented here.

%cutting due to word limit: \subsubsection{Communication protocol constraints} The simulation assumes ideal and instantaneous data exchange between participants and the LMO. Real-world constraints such as latency, packet loss and protocol-specific formatting are not modeled, which could impact both performance and vulnerability surface.

\section{Recommendations and Guidance to mitigate LEM cybersecurity}

This section presents key recommendations for researchers, industry, and policymakers based on the previously discussed issues in the article. Also, implementable guidance to help secure LEMs are provided for each LEM stakeholder.

\subsection{Recommendations for Researchers, Industry and Policymakers}

\subsubsection{For LEM researchers} Given the lack of empirical data from real LEM pilot projects, researchers should continue the development of simulation environments that replicate LEMs under cyberattack scenarios. For example, cybersecurity modeling features should be incorporated into existing open-source LEM models, such as the Open Platform for Local Energy Markets (OPLEM) platform \cite{essayeh2024oplem}. Outputs from such projects will inform and feedback improvements, such as adaptive defense mechanisms, into real LEM implementations. Engagement of collaborative innovation-policy development is also advised, where joint projects involving researchers, cybersecurity experts, policy makers, economists, and legal professionals produce insights into how different LEM models influence the exposure to cyber risks, and the corresponding alignment of needed governance. 
% \sub{Can you refer to OpLEM work here? And say we must embed cyber security features in such open-source simulators?} \Alh{addressed.}

\subsubsection{For the LEM industry (e.g. LMO platform developers)} Trading-related communications, especially those sent by LEM participants, must be authenticated. Such measures are essential to prevent spoofing, data manipulation, or unauthorized access by malicious actors or even nation-state threats \cite{Dedrick2023Assessing, Mustafa2016ALocal}. LEM platforms, devices and enabling technologies, should be developed with privacy and secure-by-design principles such as default encrypted communications and automatic patching schemes. Such features will mitigate exploitable vulnerabilities and reduce lateral movement of adversaries within LEMs \cite{Dedrick2023Assessing}. For the long term, investments in market-level intrusion detection systems (IDS) should be made. 
 
\subsubsection{For governments and regulatory bodies} To our knowledge, there are no LEM-specific cybersecurity regulations, therefore  bodies such as the UK’s Energy Networks Association (ENA) should consider developing regulations tailored for LEM cybersecurity, mentioning minimum requirements and best practices. Also, independent certification and compliance checks should be established, for example requiring all LEM-connected smart devices and platforms to undergo regular third-party security certification \cite{Dong2022Cybersecurity}.
% \cite{ena2020cyber} cutting refs

\subsection{Practical Guidance for LEM Stakeholders}

\subsubsection{For consumers and prosumers} They should recognize that insecure LEM-connected IoT devices can negatively impact LEMs. Therefore, they should change default passwords to strong passwords, enable multi-factor authentication, and install regular security updates. They should also learn to identify phishing emails and social engineering tactics. More advanced advice includes utilizing home network segmentation.

\subsubsection{For aggregators} We recommend enrolling only LEM participants with DERs that meet minimum cybersecurity standards and regulations (e.g. the UK's DER Cyber Security Connection Guidance). Once enrolled, those assets should be monitored continuously using automated tools to detect malicious devices and anomalous behavior. Finally, we advise placing emphasis on customer education by informing participants about their cybersecurity responsibilities through guidance documents, regular emails, and other communication channels. %\cite{ena2020cyber} cutting refs

\subsubsection{For LMOs} We similarly advise enforcing compliance for all LEM participants. Importantly, communication standards should incorporate strong security features (e.g. TLS) and be implemented properly to avoid implementation vulnerabilities. Maintaining a cybersecurity response plan in coordination with the DSO, TSO, and regulators is advised.

\section{Conclusion}

This work has systematically unpacked how LEMs depend on interoperable communication standards and high-wattage IoT-enabled devices, revealing attack vectors in protocols like OpenADR, IEEE 2030.5 and IEC 61850 and weaknesses in DERs and ESAs. By mapping signal flows to existing standards, analyzing key vulnerabilities, and simulating two false-data injection attacks, we demonstrated tangible distortions in local prices, demand patterns and grid operations. Our insights inform immediate measures, secure‐by‐design implementations, authentication, and participant monitoring, and longer-term needs for LEM-specific regulation and intrusion-detection. Future research should integrate wholesale-market dynamics, adaptive attacker behaviors and coordinated multi-vector scenarios to further safeguard emerging energy systems.

\section*{Acknowledgments}
This work was supported in part by a joint project between King Abdullah University of Science and Technology and the University of Warwick under Award No. RFS-OFP2023-5505.

\bibliographystyle{ieeetr}
\bibliography{Ref}

\end{document}